\documentclass[preprint,table]{revtex4-2}
\usepackage[T1]{fontenc}
\usepackage{graphicx}
\usepackage{rotating}
\usepackage{bm}        
\usepackage{amssymb}   
\usepackage{bm}
\usepackage{float}
\usepackage{tikz}
\usepackage{amsmath}
\usepackage{multirow}
\usepackage{array}
\usepackage{hyperref}
\usepackage{xcolor}
\usepackage{tabularx}
\usepackage{nicematrix}
\usepackage[normalem]{ulem}
\usepackage{booktabs}

\def\k1{k_1}
\def\k2{k_2}
\def\q1{q_1}
\def\q2{q_2}

\def\({\left (}
\def\){\right )}
\def\[{\left [}
\def\]{\right ]}
\def\gdmlf{GDML }

\def\agdmlf{AGDML }

\def\agdmlcf{AGDML(c) }

\newcommand{\beq}{\begin{equation}}
	\newcommand{\eeq}{\end{equation}}

\begin{document}
	\date{\today}
	\flushbottom \draft
	\title{Gradient domain machine learning with composite kernels: improving the accuracy of PES and force fields for large molecules}
	\author{K. Asnaashari and R. V. Krems}
	\affiliation{
		Department of Chemistry, University of British Columbia, Vancouver, B.C. V6T 1Z1, Canada
	}
	\begin{abstract}
		
		The generalization accuracy of machine learning models of potential energy surfaces (PES) and force fields (FF) for large polyatomic molecules can be improved either by increasing the number of training points or by improving the models. 
		In order to build accurate models based on expensive {\it ab initio} calculations, much of recent work has focused on the latter. In particular, it has been shown that gradient domain machine learning (GDML) models produce accurate results for high-dimensional molecular systems with a small number of {\it ab initio} calculations. The present work extends GDML to models with composite kernels built to maximize inference from a small number of molecular geometries. 
		We illustrate that GDML models can be improved by increasing the complexity of underlying kernels through a greedy search algorithm using Bayesian information criterion as the model selection metric. We show that this requires including anisotropy into kernel functions and produces models with significantly smaller generalization errors. 
		The results are presented for ethanol, uracil, malonaldehyde and aspirin. 
		For aspirin, the model with composite kernels trained by forces at 1000 randomly sampled molecular geometries produces a global 57-dimensional PES with the mean absolute accuracy 0.177 kcal/mol (61.9 cm$^{-1}$) and FFs with the mean absolute error 0.457 kcal/mol~\AA$^{-1}$. 
	\end{abstract}
	
	\maketitle
	
	\clearpage
	\newpage

	\section{Introduction}

	Accurate potential energy surfaces (PES) and force fields (FF) are required for simulations of dynamics of molecules. A major recent effort has been to develop accurate models of PES and FFs for large polyatomic molecules with accuracy of {\it ab initio} calculations. 
	As the complexity of molecules grows, it becomes increasingly difficult to produce accurate analytical fits of PES and FFs as choosing suitable functions for the parameterization becomes challenging. This problem can be addressed with machine learning (ML), as illustrated by a large body of recent work on neural network \cite{ML-for-PES2, ML-for-PES3, NNs-for-PES, NNs-for-PESa, NNs-for-PES-1a, NNs-for-PES-1b, NNs-for-PES-1c, NNs-for-PES-2, NNs-for-PES-3, NNs-for-PES-4, NNs-for-PES-5, NNs-for-PES-6, carrington, meuwly} and kernel regression \cite{meuwly, gp-1, gp-2, gp-3, jie-jpb, gp-for-PES-2, gp-for-PES-3, gp-for-PES-4, gp-for-PES-5, gp-for-PES-6, gp-for-PES-7, gp-for-PES-8, gp-for-PES-9, gp-for-PES-10, gp-for-PES-11, gp-for-PES-12, kernel-for-PES, rabitz-1, rabitz-2, rabitz-3, unke} models of PES and FFs. These ML models are generally trained by potential energies and/or forces computed with {\it ab initio} methods for different molecular geometries. 
	The accuracy of ML models generally increases with the number of training data.  However, {\it ab initio} calculations are expensive. Therefore, a significant focus of recent work has been on  building accurate ML models with as few {\it ab initio} calculations as possible \cite{general-ML, BML, ML-for-PES, gdml-1, gdml-2, sgdml, gdml-thesis}. 
	
	For problems with a small number of training points, kernel regression models have been shown to produce accurate PES and outperform NNs in some cases \cite{gp-for-PES-4}. The accuracy of kernel models largely depends on (a) the descriptors used for the input variables \cite{fchl18, fchl19}; (b) the type of model; and (c) the mathematical form of the kernel \cite{gp-for-PES-11, gp-for-PES-12, extrapolation-1, extrapolation-2, extrapolation-3, sm-1, sm-2, sm-3}. Of different model approaches, gradient domain machine learning (GDML) has so far proven to yield the most accurate results for molecules with $\sim 10 - 57$ degrees of freedom when the number of training molecular geometries is restricted to $\lesssim$ 5000  \cite{gdml-1, gdml-2, sgdml, gdml-thesis}. GDML models are trained by forces or by combinations of forces and energies to produce accurate force fields and PES. The accuracy of GDML models can be improved by building molecular symmetries into underlying kernels, which produces symmetrized  models \cite{sgdml}, hereafter denoted as sGDML. Symmetrization effectively reduces the size of the input space. However, all of the previous GDML calculations have been performed with simple, isotropic kernels that do not discriminate between input dimensions. 
	
	The effect of kernel complexity on the accuracy of PES has been explored in applications of Gaussian process (GP) regression to both low (4 atoms \cite{gp-for-PES-11}) and high (19 atoms \cite{gp-for-PES-12}) dimensional  molecules. Refs. \cite{gp-for-PES-11, gp-for-PES-12} showed that the accuracy of the GP models of PES can be enhanced by increasing the complexity of underlying kernels through a greedy algorithm combining simple mathematical functions guided by the Bayesian information criterion (BIC) as the model selection metric. This kernel selection algorithm is based on an earlier work demonstrating GP models with enhanced prediction power for pattern recognition problems \cite{extrapolation-1, extrapolation-2} and applications of GP models with composite kernels to extrapolation of properties of quantum systems in Hamiltonian parameter spaces \cite{extrapolation-3}. 
	In the present work we combine the kernel construction method of Refs. \cite{extrapolation-1, extrapolation-2, extrapolation-3, gp-for-PES-11, gp-for-PES-12} with the GDML approach to improve the accuracy of GDML models.  We demonstrate that the resulting models benefit simultaneously from the GDML formalism based on training models with forces and from the BIC-guided model selection approach.
	
	Previous GDML models were built as kernel ridge regression models with the kernel parameters determined by cross-validation using grid search \cite{gdml-1, gdml-2, sgdml}. The grid-search approach is suitable for simple, isotropic kernels that depend on a small  number of parameters. In order to take advantage of BIC, one needs to ensure that models can be trained by maximizing log marginal likelihood (LML). The first result of the present work illustrates that it is necessary to include kernel anisotropy in order to train GDML models by LML maximization. Our analysis shows that LML, and hence BIC, is not a good metric for model selection in GDML with isotropic kernels. Once the kernel anisotropy is included, however, the BIC can be used to enhance the complexity of the GDML kernels. We build composite kernels for four different molecules and illustrate the effect of kernel complexity on improving the accuracy of GDML predictions. In some instances the resulting GDML models are shown to be more accurate than the sGDML predictions.
	
	The remainder of this paper is organized as follows. The subsequent section begins with a brief summary of the notation used throughout this article and the description of the ML methods. Section III presents the results by first discussing the effect of the kernel anisotropy and then the kernel complexity. The results are presented for the PES and FFs for four molecules: ethanol (9 atoms, 21 degrees of freedom), malonaldehyde (9 atoms, 21 degrees of freedom), uracil (12 atoms, 30 degrees of freedom), and aspirin (21 atoms, 57 degrees of freedom). The accuracy of the corresponding models is compared with the previous GDML and sGDML calculations. We show that the GDML model with composite kernels produces a global 57-dimensional PES for aspirin with the mean absolute accuracy 0.177 kcal/mol (61.9 cm$^{-1}$) and FF with the mean absolute accuracy 0.457 kcal/mol~\AA$^{-1}$ when trained by 1000 randomly sampled molecular geometries.  
	
	\newpage
	
	\section{Methods}

	\subsection{Model abbreviations}
	
	We present and discuss results for several ML models of PES and FFs. Table I lists the acronyms used throughout this work.  
	The accuracy of the models is quantified by the mean absolute error (MAE):
	\begin{eqnarray}
		\textrm{MAE} = \frac{1}{n} \sum_{i=1}^{n} | y_i - \hat{f}(\bm x_i)|
		\label{MAE}
	\end{eqnarray}
	and the root mean squared error (RMSE):
	\begin{eqnarray}
		\textrm{RMSE} = \sqrt{\frac{1}{n} \sum_{i=1}^{n} \left( y_i - \hat{f}(\bm x_i) \right)^2}
		\label{RMSE}
	\end{eqnarray}
	evaluated on a hold-out set of $n$ potential energies or force fields, randomly sampled from the entire configuration space and not used for training the models. One exception is the training energy error, hereafter denoted `Train E', that is computed as the RMSE with the energy points in the training set. 
	In Eq. (\ref{MAE}), $\bm x_i$ is a $p$-dimensional vector specifying the positions of atoms in a molecule, 
	$y_i$ represents the potential energy or forces experienced by each individual atom at $\bm x_i$, and $\hat f(\bm x_i)$ denotes the prediction of the ML model at $\bm x_i$. 
	We build models $\hat{f}_E$ for energy and $\hat{f}_F$ for forces. 
	
	\begin{table}[htbp]
		\begin{NiceTabular}{ll} \hline 
			Abbreviation\quad & Meaning \\ \hline
			GDML & Gradient domain machine learning \\ 
			sGDML & GDML with symmetrized kernels \\ 
			\agdmlf & GDML with simple anisotropic kernels \\
			\agdmlcf & GDML with composite anisotropic kernels \\  
			MAE  & Mean absolute error \\
			RMSE  & Root mean squared error \\
			LML & Log marginal likelihood \\
			Train E & Training energy error (RMSE) \\
			\hline
		\end{NiceTabular}
		\caption{Abbreviations used in this work.}
		\label{table:abbrev}
	\end{table}
	
	\subsection{GDML}
	
	GDML models explicitly construct an energy-conserving force field by implementing the relation between the energy of the molecule and the forces acting on each atom as an a priori condition of the model \cite{gdml-1}:
	\begin{equation}\label{eq:ef}
		\hat{f}_F(\bm x) = -\nabla\hat{f}_E(\bm x), 
	\end{equation}
	where for a molecular system of $N$ atoms, $\bm x \in\chi^{3N}$ represents the coordinates of the atoms, $\hat{f}_E(\bm x):\chi^{3N}\rightarrow \mathbb{R}$ is an estimator of the energy, $\hat{f}_F(\bm x):\chi^{3N}\rightarrow \mathbb{R}^{3N}$ is an estimator of the forces, and $\nabla$ is the gradient operator. If we consider the energy estimator as a realization of a GP \eqref{eq:e_gp}, since $\nabla$ is a linear operator, the estimator of forces will also be a realization of a GP \eqref{eq:f_gp}: 
	\begin{equation}\label{eq:e_gp}
		\hat{f}_E\sim {\rm GP}\left[\mu(\bm x), k(\bm x,\bm x')\right]
	\end{equation}
	\begin{equation}\label{eq:f_gp}
		\hat{f}_F\sim {\rm GP}\left[-\nabla\mu(\bm x), \nabla_{\bm x}k(\bm x,\bm x')\nabla_{\bm x'}^T\right]
	\end{equation}	
	where $\mu(\bm x):\chi^{3N}\rightarrow \mathbb{R}$ and $k(\bm x,\bm x'):\chi^{3N}\times\chi^{3N}\rightarrow \mathbb{R}$ are the mean and covariance functions of the GP.
	One can also model both forces and energies as a single GP through the methodology described by Solak et al.\cite{derivative_gp}:  
	\begin{equation}\label{eq:fe_gp}
		\hat{f}_{FE}\sim {\rm GP}\left[\begin{bmatrix}
			\nabla\mu(\bm x) \\
			\mu(\bm x)
		\end{bmatrix}, \begin{bmatrix}
			\nabla_{\bm x}k(\bm x,\bm x')\nabla_{\bm x'}^T & \nabla_{\bm x}k(\bm x,\bm x') \\
			k(\bm x,\bm x')\nabla_{\bm x'}^T & k(\bm x,\bm x')
		\end{bmatrix}\right].
	\end{equation}
	The models given by Eq. \eqref{eq:fe_gp} require both forces and energies for each molecular configuration in the training data, while the models in Eq. \eqref{eq:f_gp} require only the mean of the energies in the training set. 
	Previous studies have shown that these hybrid models overfit the energies at the cost of the forcefield accuracy \cite{gdml-thesis}. We observed that both types of models when trained using log marginal likelihood yield very similar predictions for the same training sets. Therefore, in what follows, we use models trained by forces only, i.e. models given by Eq.  \eqref{eq:f_gp}. 
	
	Equation \eqref{eq:f_gp} describes a multi-output GP which predicts the vector components of forces for each atom in a molecule. The covariance function of the derivative of a GP is the second derivative of the original kernel function and $\nabla_{\bm x}k\nabla_{\bm x'}^T = \text{Hess}_{\bm x} (k) = k_H(\bm x ,\bm x') \in \mathbb{R}^{3N\times 3N}$ for stationary kernels. The posterior mean of the model of forces is
	\begin{equation}\label{eq:gdmlf}
		\hat{f}_F({\bf X^*}) = \mathbf{\alpha}k_H({\bf X^*},{\bf X})^T = \sum_i^M\sum_j^{3N}(\alpha_i)_j\frac{\partial}{\partial x_j}\nabla k({\bf X^*},\bm x_i)
	\end{equation}
	where ${\bf X} \in \mathbb{R}^{M\times 3N}$ are the training geometries (3 coordinates for each of $N$ atoms of $M$ training geometries), ${\bf X^*} \in \mathbb{R}^{M'\times 3N}$ are the molecular geometries corresponding to $M'$ evaluation points in the configuration space, $k_H({\bf X^*}, {\bf X}) \in \mathbb{R}^{3M'N\times 3MN}$ is the kernel matrix coupling the training and evaluation geometries, $\frac{\partial}{\partial x_j}$ is a partial derivative with respect to the $j$th dimension of $\bm x_i$ and $\bm \alpha \in \mathbb{R}^{3MN}$ is defined as: 
	\begin{equation}
		\bm \alpha \equiv (K_H+\lambda\mathbb{I})^{-1} \bm y_F
	\end{equation}
	where $K_H = k_H({\bf X},{\bf X}) \in \mathbb{R}^{3NM\times 3NM}$, $\bm y_F \in \mathbb{R}^{3NM}$ are the components of the forces in the training set flattened into a one-dimensional vector and $\lambda$ is the parameter related to the variance of the Gaussian noise of the training data (equivalent to the regularization parameter in kernel ridge regression). While the training data in this work are noiseless, the parameter $\lambda$ will be shown to play an important role due to overcompletness of the descriptors (see more details in Section \ref{anisotropy}). 
	
	By integrating Eq. \eqref{eq:gdmlf}, one obtains the model of the PES: 
	\begin{equation}
		\hat{f}_E({\bf X^*}) = \bm \alpha k_G({\bf X^*},{\bf X})^T = \sum_i^M\sum_j^{3N}(\alpha_i)_j\frac{\partial}{\partial x_j}k({\bf X^*},\bm x_i) + c
	\end{equation}
	where $k_G$ is defined as the gradient of the kernel function with respect to the input dimensions $k_G({\bf X^*},{\bf X}) \in \mathbb{R}^{M'\times 3MN}$ and $c$ is the integration constant which can be calculated as 
	\begin{equation}\label{eq:gdmlf_c}
		c = \frac{1}{M}\sum_i^M \left[E_i + \hat{f}_E(\bm x_i)\right],
	\end{equation}
	where $E_i$ is the energy of the $i$th geometry of the training set.
	
	Given the kernel matrix and a value of $\lambda$,  the logarithm of marginal likelihood (LML) can be calculated for these models as follows:
	\begin{equation}\label{eq:gdmlf_mll}
		\log p(\bm y_F |{\bf X}) = -\frac{1}{2}\bm y_F^T \left(K_H+\lambda\mathbb{I}\right)^{-1} \bm y_F - \frac{1}{2}\log |K_H+\lambda\mathbb{I}| - \frac{M}{2}\log 2\pi
	\end{equation}
	
	Note again that in this formulation the model does not use energy points directly so energies are not used for building the PES. Energy predictions are determined by the individual energy points indirectly, through forces, and through the mean of the energies (first term in Eq. \eqref{eq:gdmlf_c}). 
	LML for such models is consequently independent of the energies and is written in terms of forces $\bm y_F$ only.
	
	\subsection{Kernel functions in GDML}
	
	The GDML method described above can, in principle, be used with any doubly differentiable stationary kernel function. Some examples of such kernel functions commonly used for kernel regression models include 
	Mat\'ern functions of order $\geq 5/2$, radial basis functions (RBF) or rational quadratic (RQ) functions \cite{extrapolation-2}: 
	\begin{equation}
		\textrm{RBF:}~~~	k(\bm x,\bm x') = \sigma \exp\left(-\frac{1}{2}r^2(\bm x,\bm x')\right) \label{eq:rbf}
	\end{equation}
	\begin{equation}
		\textrm{Mat\'ern 5/2:}~~~		k(\bm x,\bm x') = \sigma \left(1+\sqrt{5}r(\bm x,\bm x')+\frac{5}{3}r^2(\bm x,\bm x')\right)\exp\left(-\sqrt{5}r(\bm x,\bm x')\right) \label{eq:mat52}
	\end{equation}
	\begin{equation}
		\textrm{RQ:}~~~	k(\bm x,\bm x') = \sigma \left(1+\frac{1}{2\alpha}r^2(\bm x,\bm x')\right)^{-\alpha} \label{eq:ratquad}
	\end{equation}
	where $r^2(\bm x,\bm x') = (\bm x-\bm x')^T\times \Lambda \times (\bm x-\bm x')$ and $\Lambda$ is a diagonal matrix. For isotropic kernels, $\Lambda = l\times\mathbb{I}$ with $l$ being a positive scalar and $\mathbb{I}$ an identity matrix.
	To the best of our knowledge, previous studies and extensions of the GDML models \cite{gdml-1, gdml-2, sgdml} use the isotropic Mat\'ern 5/2 kernel function. 
	In the previous studies, $\sigma$ in Eq. (\ref{eq:mat52}) was set to 1 and the single value of $l$ was determined by grid search using cross-validation. 
	
	The main goal of the present work is to extend the previous GDML work to include composite kernel functions built using the methodology developed by Duvenaud et al. \cite{extrapolation-1, extrapolation-2, extrapolation-3}.
	We show below that this requires allowing for kernel anisotropy. 
	For anisotropic kernels, $\Lambda$ has a free parameter for each dimension of $\bm x$. 
	Given the large number of descriptor dimensions for the molecules considered here (up to 57 degrees of freedom for aspirin, translating into 210 descriptor dimensions using the descriptors discussed below), allowing for kernel anisotropy makes grid search of kernel parameters impossible. This, however, does not present a problem when models are trained by LML maximization.  
	
	\subsubsection{Simple kernels}
	
	We refer to kernels given by Eqs. (\ref{eq:rbf}) -- (\ref{eq:ratquad})  as simple, whereas any linear combination of different simple kernels is hereafter referred to as a composite kernel function.  
	For reasons described in the previous sections, we limit simple kernels to doubly differentiable stationary kernel functions and specifically to the set of three kernel functions given by  Eqs. (\ref{eq:rbf}) -- (\ref{eq:ratquad}). 
	We consider models with simple kernels that are either isotropic or anisotropic. As specified in Table \ref{table:abbrev}, GDML models with simple anisotropic kernel functions will be denoted by AGDML.
	A model for aspirin based on a simple anisotropic kernel given by Eq. (\ref{eq:mat52}) has $\approx 210$ trainable parameters. Note that the number of trainable parameters increases substantially as composite kernels are formed from simple kernels.

	\subsubsection{Composite kernels} \label{composite_kernels}
	
	As shown previously, composite kernel functions cannot be constructed as random combinations of simple kernels \cite{gp-quant}. Instead, the kernel construction algorithm should be guided by a model selection metric. 
	In this work, composite kernel functions are built using the methodology of Ref. \cite{extrapolation-2} that involves a greedy search algorithm with the BIC as the model selection metric. The BIC is defined as follows: 
	\begin{equation}
		\textrm{BIC} = \text{max(LML)} - \frac{\gamma}{2}\log(M)
	\end{equation}
	where $\gamma$ is the number of free parameters in the model and $M$ is the number of training geometries. 
	This kernel selection algorithm can be viewed as a search tree with the following steps: 
	\begin{enumerate}
		\item Build models using simple kernel functions;
		\item Evaluate the BIC for each model;
		\item Select the model with the largest value of BIC as a base model;
		\item Add and multiply the kernel of the base model with each of  the simple kernel functions used in step 1 to create new kernel functions and train the models with the new composite kernel functions;
		\item Repeat steps 2 - 4 until the improvements in test (validation) error become negligible or the error begins to rise due to overfitting.
	\end{enumerate}
	Each iteration of steps 2 - 4 creates a new layer of the search tree. As the algorithm progresses to deeper layers, the complexity of the optimal kernel function increases. 	
	
	This algorithm has been shown to increase the accuracy of PES of molecules using energy-based models \cite{gp-for-PES-11, gp-for-PES-12}. 
	Here, we use this methodology with anisotropic kernel functions to build GDML models based on forces. Since we work with gradients and Hessians of the kernel functions, only linear combinations of kernel functions is considered in step 4 to simplify the kernel search in the present work. As specified in Table \ref{table:abbrev}, GDML models with anisotropic composite kernels are referred to \agdmlcf.
	
	\section{Results}
	
	All results in this work use the MD17 dataset \cite{ML-for-PES2}. The molecules considered include ethanol, malonaldehyde, uracil, and aspirin. The energies and forces in the dataset were computed using the PBE + vdW-TS electronic structure method \cite{md17-1, md17-2}. Table \ref{table:datasets} displays the number of configurations and the range of energies and forces in the datasets. We do not consider some of the molecules in MD17 dataset (benzene, naphthalene, and salicylic acid). These molecules are rigid and as a result their potential energy surface and force fields are quite simple to learn. Due to their simplicity, these molecules are well modeled by basic GDML (as evidenced by the fact that extension to sGDML does not result in significant accuracy improvements \cite{sgdml}) and are not expected to benefit significantly from AGDML.
	
	\begin{table}[htbp]
		\begin{NiceTabular}{cccccc} \hline 
			\multirow{2}{*}{Molecule} & \multirow{2}{*}{\# of configurations} & Minimum energy & Maximum energy & Minimum force & Maximum force \\ 
			& & (kcal/mol) & (kcal/mol) & (kcal/mol~\AA$^{-1}$) & (kcal/mol~\AA$^{-1}$) \\ \hline
			Ethanol & 555092 & -97208.4 & -97171.8 & -211.1 & 220.9 \\ 
			Malonaldehyde & 993237 & -167514.2 & -167470.4 & -286.0 & 284.6 \\ 
			Uracil & 133770 & -260120.6 & -260080.8 & -237.3 & 239.2 \\
			Aspirin & 211762 & -406757.6 & -406702.3 & -210.4 & 213.4 \\  
			\hline
		\end{NiceTabular}
		\caption{Number of configurations and the range of energies and forces for each of the four molecules in the MD17 dataset \cite{ML-for-PES2} used in this work.}
		\label{table:datasets}
	\end{table} 
	
	\subsection{Effects of kernel anisotropy} \label{anisotropy}
	
	We begin by considering the effects of kernel anisotropy in models with simple kernels. The kernel anisotropy is included 
	by increasing the number of trainable parameters of the kernel function to match the number of the input variables. 
	Following previous work \cite{gdml-1, gdml-thesis}, we use the following descriptor of molecules based on the Coulomb matrix: 
	\begin{equation}
		D_{ij} = \begin{cases}
			\lVert \bm r_i - \bm r_j \rVert^{-1} & \text{for } i>j \\
			0 & \text{for } i\leq j
		\end{cases} \label{eq:desc}
	\end{equation}
	where $\bm r_i$ is the vector of coordinates of the $i$th atom in the molecule.
	Coulomb matrices benefit from the roto-translational invariance of the molecular systems. For a molecule with $N$ atoms, there are $3N$ input dimensions (atom coordinates) and $N(N-1)/2$ descriptor dimensions. As each molecule has $3N-6$ independent degrees of freedom, the descriptor dimensions are not independent for $N>4$.
	
	Given a descriptor function $D(\bm x)$, the kernel function $k_D(\bm x ,\bm x' ) = k(D(\bm x),D(\bm x'))$ and the Hessian and gradient of the kernel function w.r.t. the input dimensions (as opposed to the descriptor dimensions) become:
	\begin{equation}
		k_H = J_D(\nabla_Dk_D\nabla_{D'}^T)J_{D'}^T
	\end{equation}
	\begin{equation}
		k_G = J_D\nabla_Dk_D
	\end{equation}
	where $J_D$ is the Jacobian of the descriptor function w.r.t. the input dimensions. With $D(\bm x)$ as inputs, anisotropic kernels have ${N(N-1)}/{2}$ parameters with $N$ being the number of atoms. Anisotropic kernels thus built are over-parameterized for $N>4$. 
	
\begin{figure}[htbp]
		\begin{tabular}{cc}
			\includegraphics{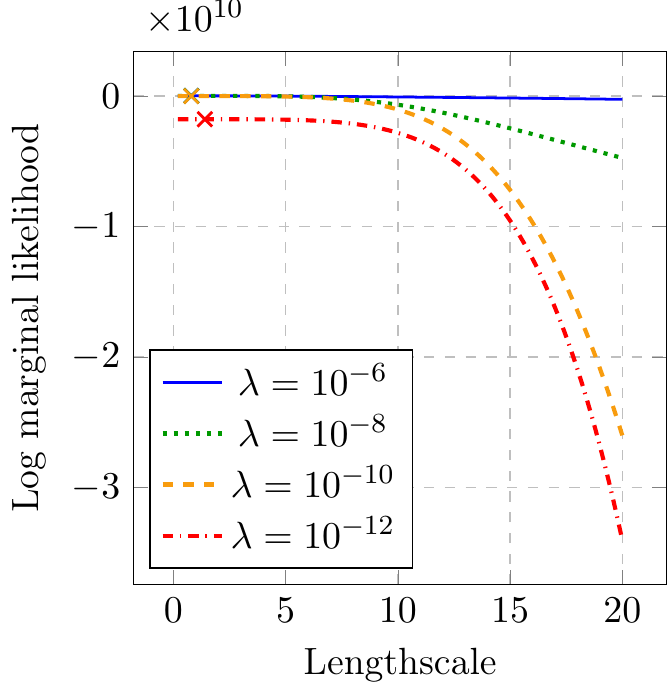} & \includegraphics{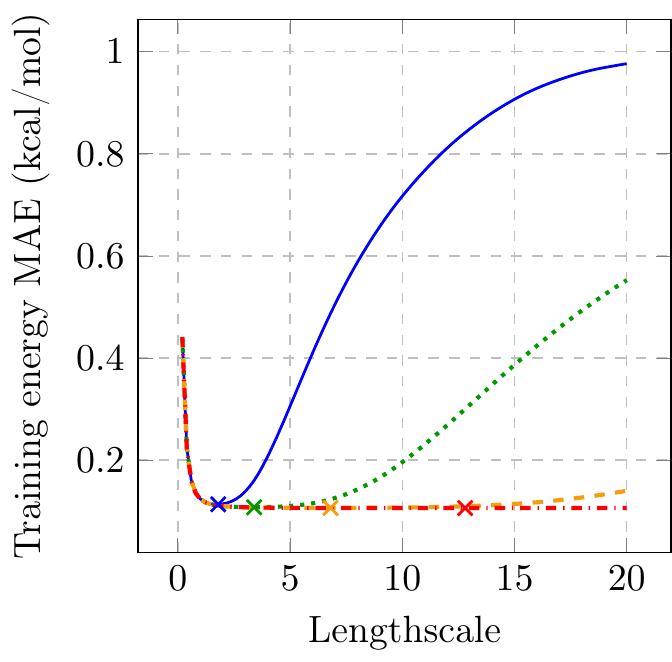} \\
			\includegraphics{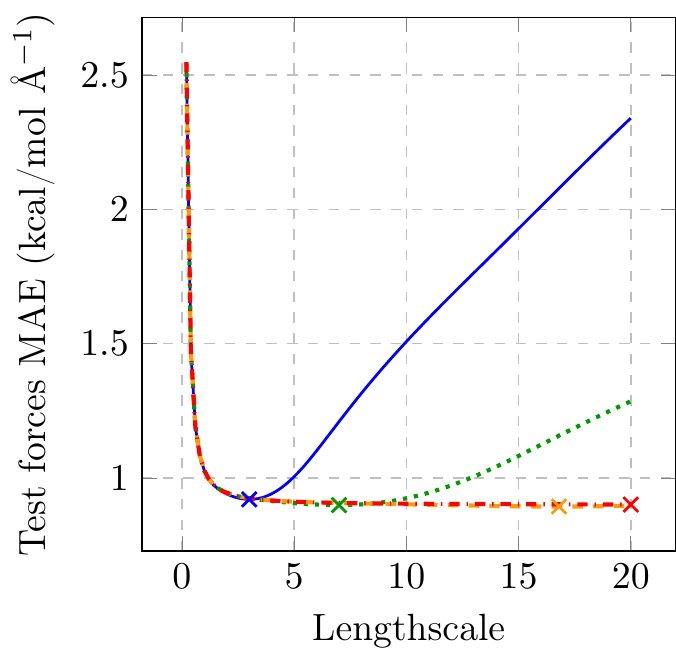} & \includegraphics{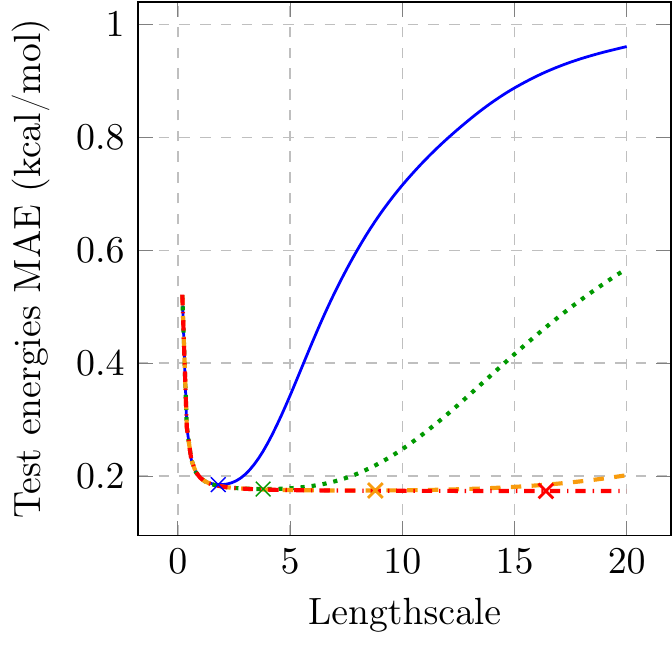}
		\end{tabular}
		\caption{\gdmlf models using the Mat\'ern 5/2 kernel with $l \in [0.2, 20]$ and $\sigma = 1$ trained on 800 and tested on 2000 molecular geometries for ethanol. The maxima of log marginal likelihood and minima of MAEs for each model are marked with an x. The different curves correspond to different values of $\lambda$: $\lambda = 10^{-6}$ (blue), $10^{-8}$ (green), $10^{-10}$ (orange), $10^{-12}$ (red).}
		\label{fig:mat52_800_noE}
	\end{figure}

	Kernel models with a large number of parameters cannot be trained by grid search. 	An alternative to grid search of kernel parameters in kernel ridge regression is maximization of log marginal likelihood in Gaussian process regression. 
	Therefore, we first consider the possibility of building GDML models with Coulomb matrix descriptors by maximizing LML. We begin by analyzing the models with 
	the isotropic Mat\'ern 5/2  kernel at different values of $\lambda$ over a range of lengthscales $l$ in Eq. \eqref{eq:mat52} for the molecule ethanol. We sample energies and forces at 800 molecular geometries to generate a training set and at 2000 geometries to generate a hold-out test set. 	
	Note that the 800 energies from the training set are not used to train the GDML models (only their mean is used to determine the constant $c$ in equation \eqref{eq:gdmlf_c}), as all models in the present work are trained by forces only. Nevertheless, we refer to these energies as the training energies, as they correspond to the molecular geometries in the training set. 
	Figure \ref{fig:mat52_800_noE} displays the log marginal likelihood, MAEs  for test energies and test forces and MAEs for training energies. 
	MAEs for training forces are not displayed as the errors of GPs on data used for training are insignificant. Energy values at the training geometries are, however, inferred from the forces and the mean of the training energies, which makes the training energy MAEs for the \gdmlf models in this work similar to the test energy MAEs.
	
\begin{figure}[htbp]
		\begin{tabular}{cc}
			\includegraphics{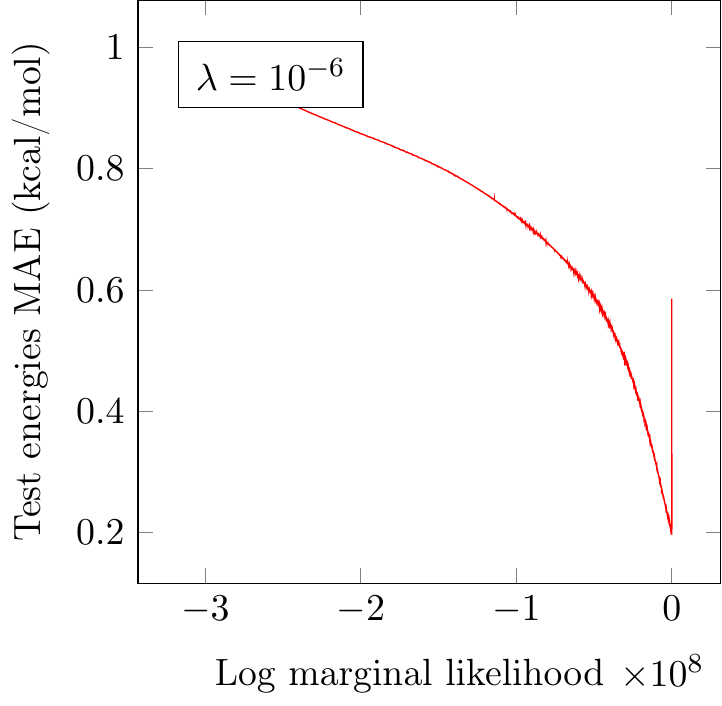} & \includegraphics{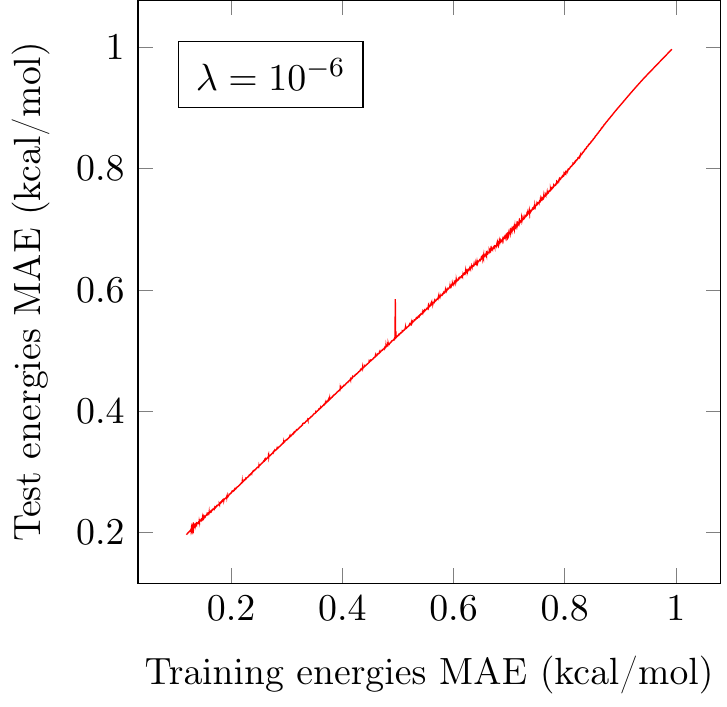} \\
			\includegraphics{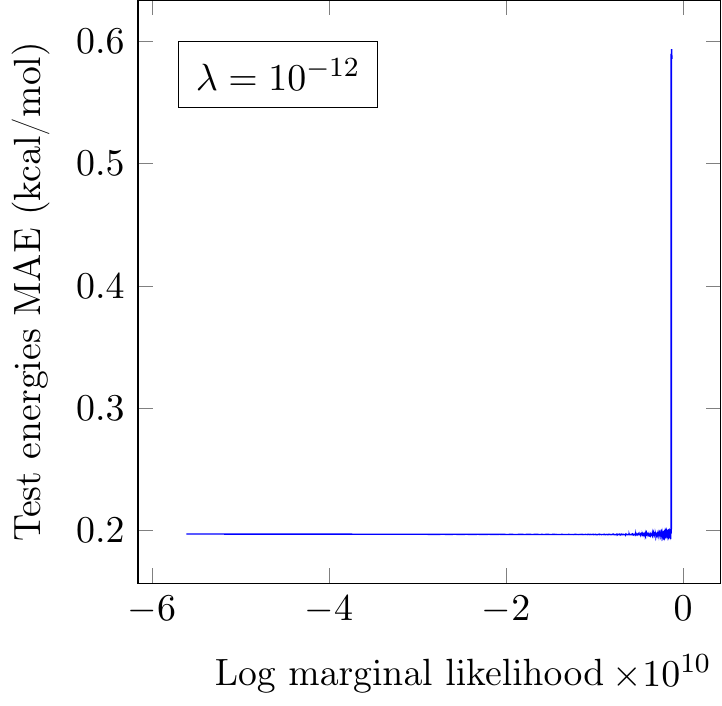} & \includegraphics{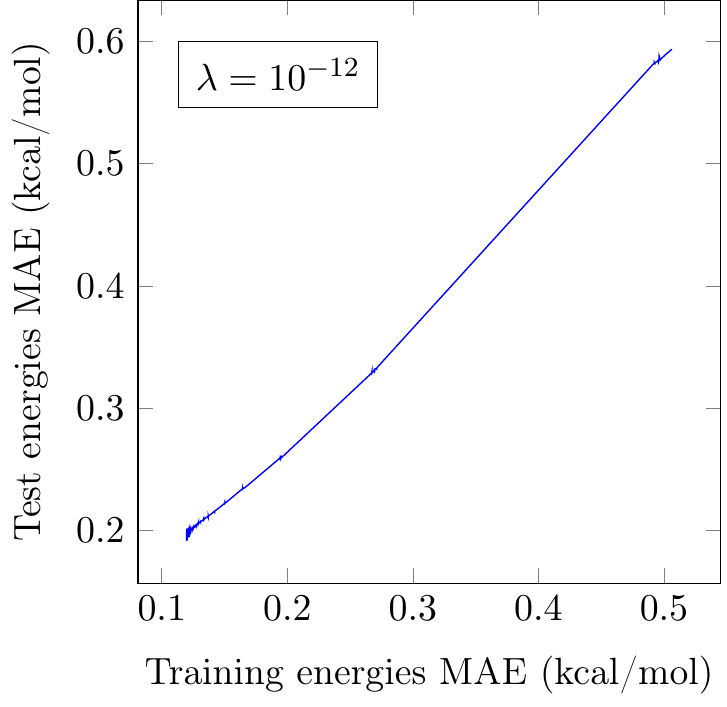}
		\end{tabular}
		\caption{Correlations between log marginal likelihood and test energy errors (left panels) and between training energy errors and test energy errors (right panels) of \gdmlf models with a Mat\'ern 5/2 kernel with $l \in [0.2, 20]$, $\sigma \in [0.5, 10]$, and  $\lambda = 10^{-6}$ (upper panels) and $\lambda = 10^{-12}$ (lower panels) trained on 600 and tested on 2000 molecular geometries for ethanol.}
		\label{fig:mat52_out_600_noE}
	\end{figure}

	Figures \ref{fig:mat52_800_noE} and \ref{fig:mat52_out_600_noE} show that LML cannot be used as a unique model metric for problems with an  isotropic Mat\'ern 5/2 kernel \eqref{eq:mat52}, as the lowest values of test errors do not correspond to the largest values of LML. 
	The results in Figure  \ref{fig:mat52_800_noE} are obtained with the value $\sigma = 1$, as in previous work \cite{gdml-1, gdml-thesis}. 
	Figure \ref{fig:mat52_out_600_noE} explores the $(\sigma, l)$ parameter space of the isotropic Mat\'ern 5/2 kernel \eqref{eq:mat52}, with $\sigma \in [0.5, 10]$ and $l \in [0.2, 20]$. The calculations are performed for several fixed values of the regularization parameter $\lambda$, as indicated in the figures. 
	
	We observe that lower values of $\lambda$ yield lower error values on test data, as should be expected for noiseless problems. However, 
	the LML values for models with small $\lambda$ are very large and negative which reflects numerical instabilities due to inversion of the unregularized kernel matrix. 
	These numerical instabilities are a consequence of the over-parametrization (i.e. the number of input variables $D(\bm x)$ is greater than the number of independent variables). This was observed in Ref. \cite{gdml-thesis}.  
	We find similar trends for models with the isotropic RBF and RQ kernels. 
	
	Figures \ref{fig:mat52_800_noE} and \ref{fig:mat52_out_600_noE}  also show that the training energy MAE correlates well with the test energy MAE. Since the energies from the training set are not used for training the GDML models, one can -- in principle -- use the MAE calculated over energies in the training set for selecting the kernel parameters. 
	To compare the efficacy of LML and training energy error for building force-based GDML models with simple kernels, we summarize the most accurate models in Table \ref{table:simple_iso}. Models labeled `Train E' are trained by minimizing the training energy RMSE (since MAE has a non-continuous gradient) using a gradient-based optimizer. The results show that models based on training energy error minimization are more accurate than those trained by LML maximization. Table \ref{table:simple_iso} also illustrates that the GDML models with simple kernels are sensitive to the choice of the kernel function. 
	
	The results are different for anisotropic kernels. As illustrated in Table \ref{table:simple_aniso}, LML for models with anisotropic kernels becomes the best model metric. Maximization of LML produces the most accurate results. In contrast, RMSE computed over training energies is no longer a good metric. Table \ref{table:simple_aniso} also illustrates that the model accuracy is sensitive to the choice of the anisotropic kernel function. Thus, the RQ anisotropic kernel yields significantly better results for both PES and FFs than the models with the  Mat\'ern 5/2 kernel. 
	Finally, a comparison of the results in  Tables \ref{table:simple_iso} and \ref{table:simple_aniso} shows that allowing for anisotropy in simple kernels leads to a significant improvement of the models. 
	
	We conclude that LML cannot be used for training GDML models with simple isotropic kernels and Coulomb matrix descriptors.  The results of Figures \ref{fig:mat52_800_noE} and \ref{fig:mat52_out_600_noE}  and Table \ref{table:simple_iso} suggest that 
	RMSE over training energies can, potentially, be used to train such  models.  However, we have found that this is prone to overfitting for problems with anisotropic kernels, making training energy RMSE not suitable for training composite GDML models. As illustrated by Table \ref{table:simple_aniso}, LML can be used for training GMDL models with anisotropic kernels. Since BIC is closely related to LML, one cannot use BIC as a model selection metric for GDML models with isotropic kernels. Kernel anisotropy is, thus, key to the kernel improvement method used in the present work.
	
	\subsection{Effect of kernel complexity}
	
\begin{figure}[htbp]
		\begin{tabular}{lcc}
			\includegraphics{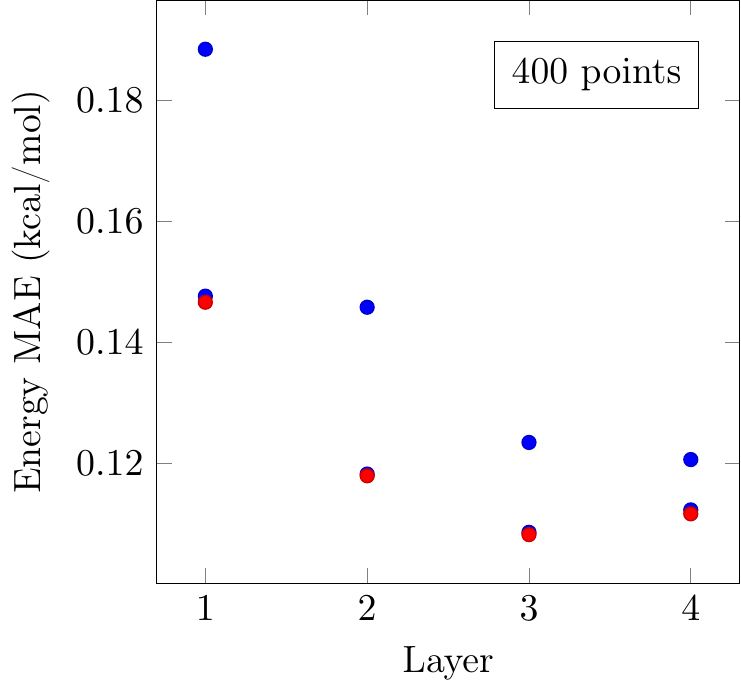} & \includegraphics{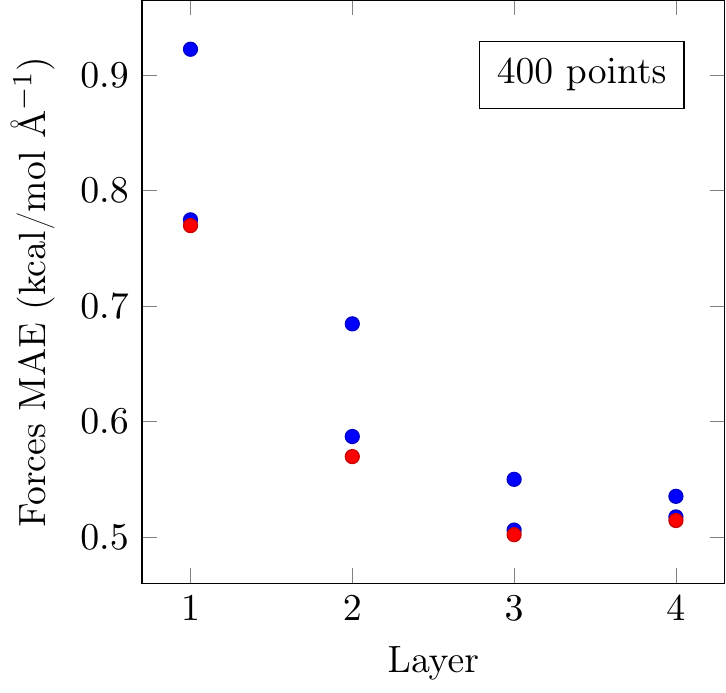} \\
			\includegraphics{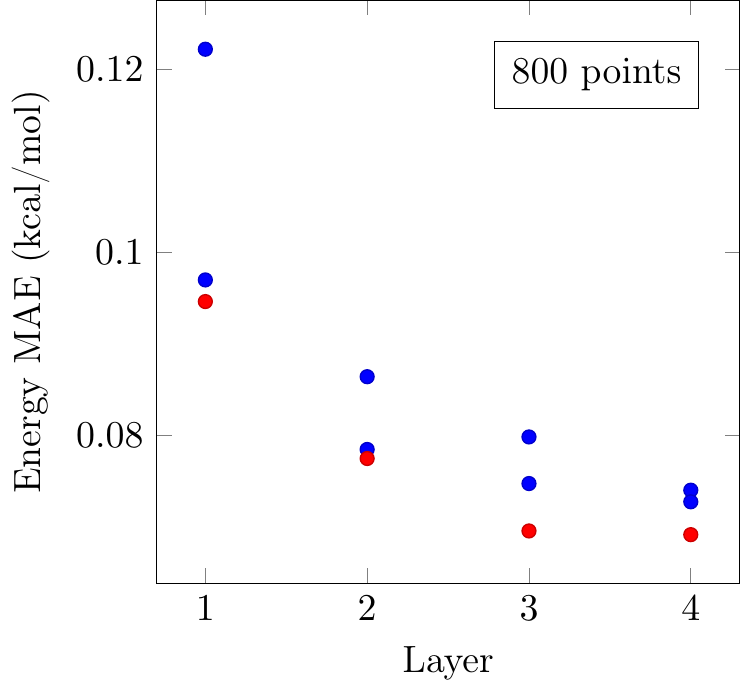} & \includegraphics{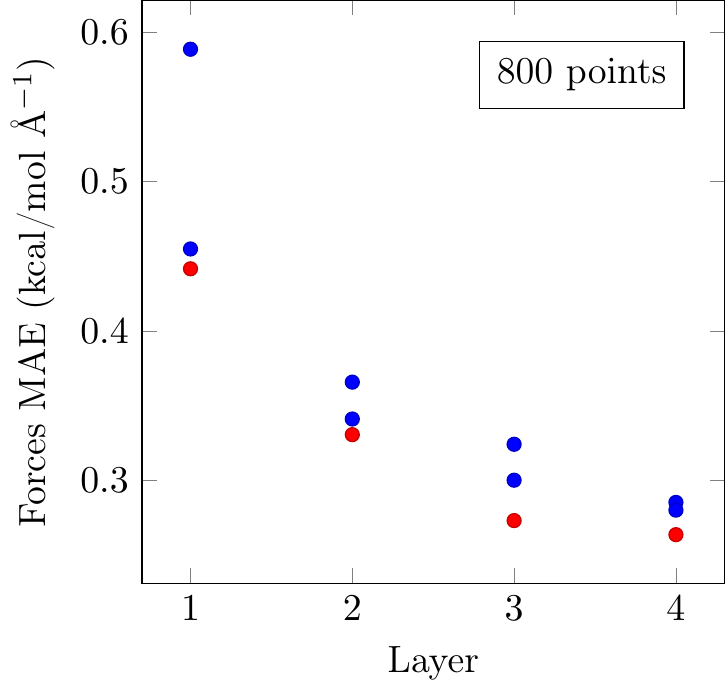} \\
		\end{tabular}
		\caption{\agdmlcf models for ethanol trained on 400 (upper panels) and 800 (lower panels) molecular geometries by LML maximization and tested on 2000 molecular geometries. 
			Each layer indicates an increase in the number of simple kernel functions in the composite kernel function. The red dots label the models with the maximum value of BIC.}
		\label{fig:comp_aniso_eth_noE_48}
	\end{figure}

\begin{figure}[htbp]
		\begin{tabular}{cc}
			\includegraphics{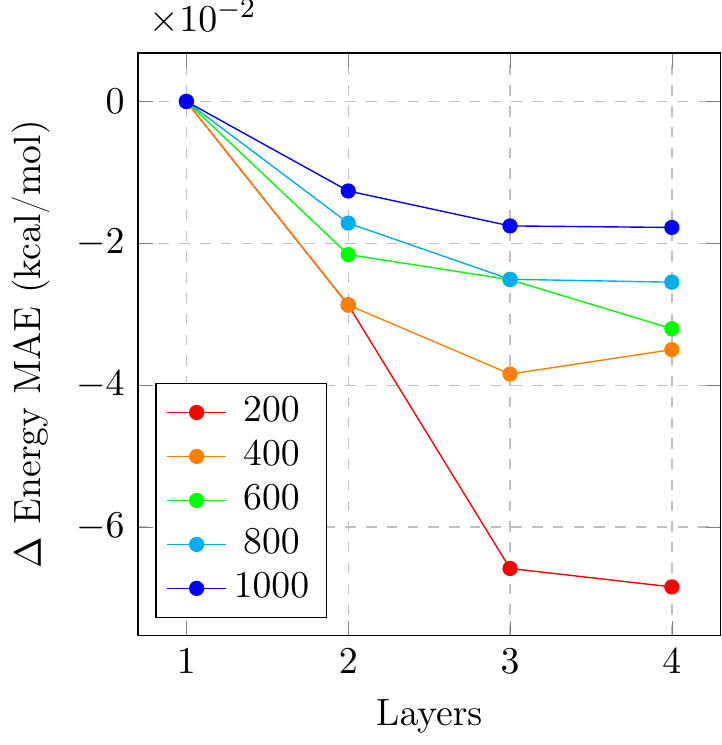} & \includegraphics{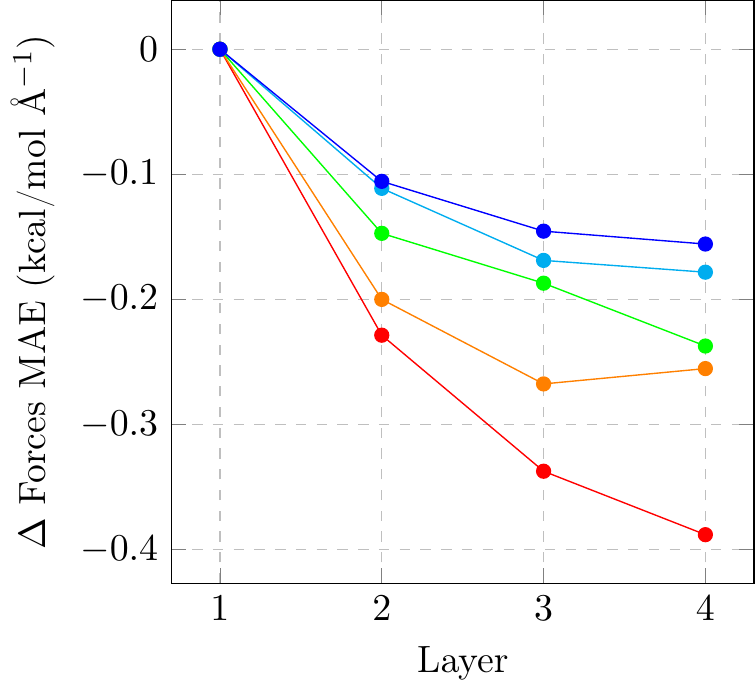} \\
		\end{tabular}
		\caption{Improvement in the test energy and forces MAEs of \agdmlcf models with increasing kernel complexity for ethanol trained with $n = 200, 400, 600, 800, 1000$ molecular geometries. The errors are computed with 2000 molecular geometries.}
		\label{fig:comp_aniso_improv_eth}
	\end{figure}

	We use simple anisotropic kernels as the basis for the kernel construction algorithm described in Section \ref{composite_kernels}. Figures \ref{fig:comp_aniso_eth_noE_48} and \ref{fig:comp_aniso_improv_eth}  illustrate the improvement of the \agdmlcf models of both energies and forces for several sets of training geometries as the kernel complexity increases. 
	Figure \ref{fig:comp_aniso_eth_noE_48} demonstrates that the maximum values of BIC correspond to the best model at each layer (iteration) of the kernel construction algorithm. 
	
\begin{figure}[htbp]
		\begin{tabular}{cc}
			\includegraphics{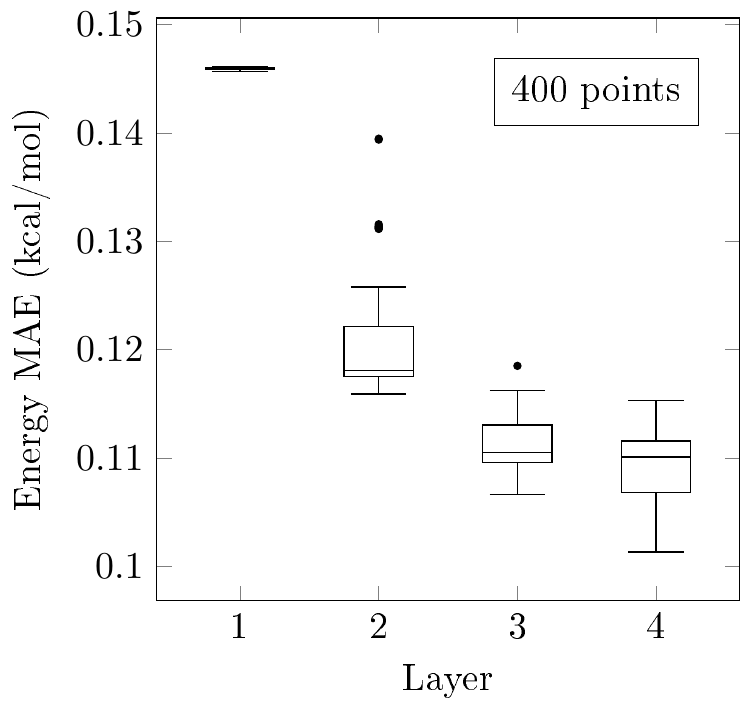} & \includegraphics{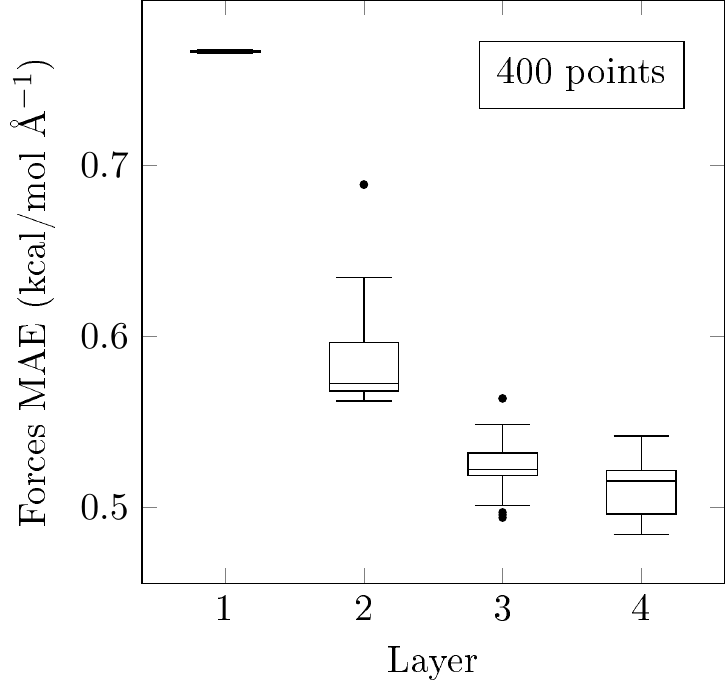} \\
			\includegraphics{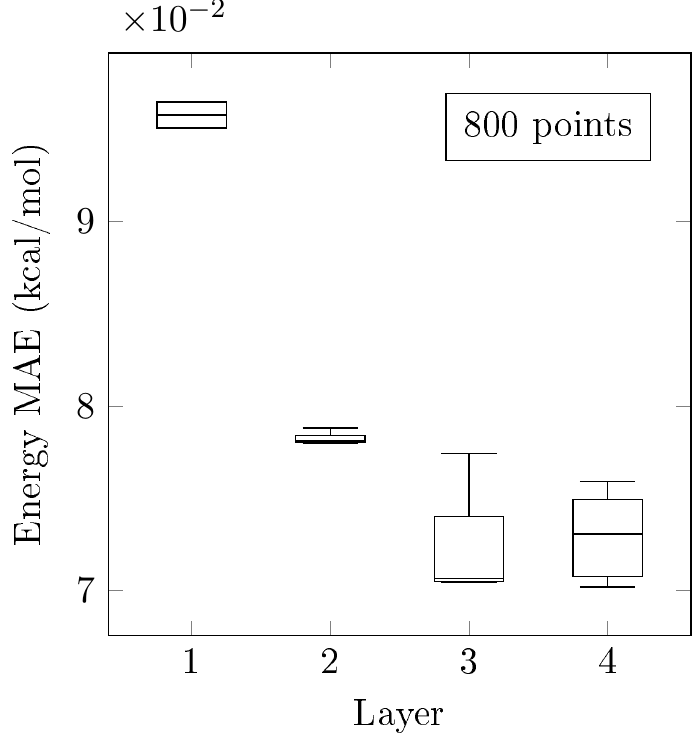} & \includegraphics{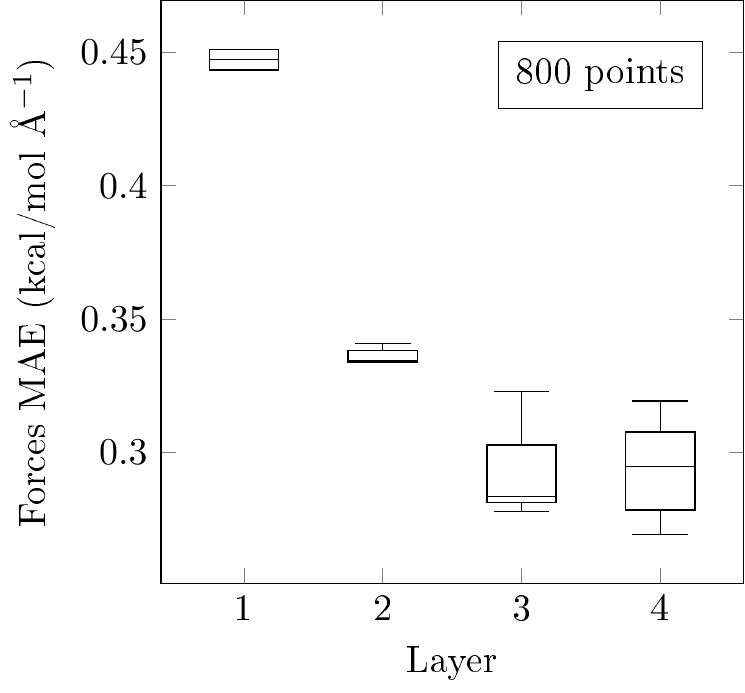} \\
		\end{tabular}
		\caption{
			Test energy and forces MAEs for \agdmlcf models for ethanol trained on 400 (upper panels) and 800 (lower panels) molecular geometries by LML maximization and tested on 2000 molecular geometries. Each layer indicates an increase in the number of simple kernel functions in the composite kernel function. The box plots show the distribution of the trained models with all the kernel hyperparameters initialized randomly (40 times for each of the 400 point models and 10 times for each of the 800 point models) from a log-normal distribution where $\log(X) \sim {\cal N}(\mu=0.7, \sigma=0.2)$.}
		\label{fig:comp_aniso_eth_dist}
	\end{figure}

	In order to evaluate the effectiveness of the training method based on LML maximization for increasingly more complex kernels, we initialized the kernels with the highest BIC of each layer for 400 and 800 training geometries of ethanol (red dots in figures \ref{fig:comp_aniso_eth_noE_48}) with random parameters from a log-normal distribution where $\log(X) \sim {\cal N}(\mu=0.7, \sigma=0.2)$. We found that this distribution of initial hyperparameters produces good results. The corresponding distributions of energy and force MAEs are shown in figure \ref{fig:comp_aniso_eth_dist} for 40 initializations on the 400 point models and 10 initializations on the 800 point models.
	
	Figures \ref{fig:comp_aniso_eth_noE_48} and \ref{fig:comp_aniso_eth_dist} also show that the accuracy of the models saturates at some iteration of the BIC-driven kernel composition algorithm. Increasing the kernel complexity from layer 3 to layer 4 results in a decrease of accuracy in figure \ref{fig:comp_aniso_eth_noE_48} on 400 points. Each iteration of the composition algorithm adds $\sim36$ kernel parameters so the number of parameters can become comparable to the number of training geometries as the complexity of the kernel function is increased. As a result, the models can become over-parameterized when trained with a small number of training geometries. This limits the number of layers of kernel complexity that offer a significant improvement of model accuracy. We observe that the model error generally saturates or, in some cases, increases due to overfitting, as the number of kernel layers is further increased. 
	
\begin{figure}[htbp]
		\begin{tabular}{cc}
			\includegraphics{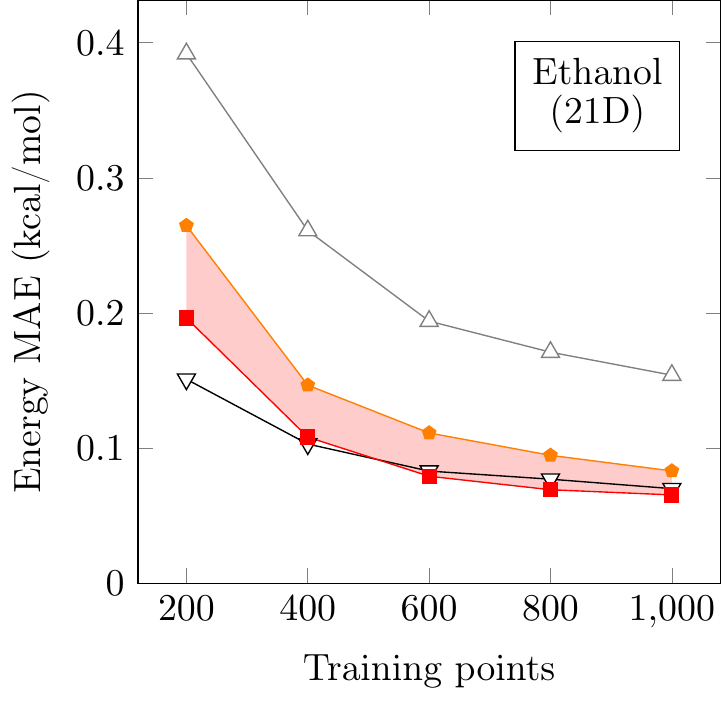} & \includegraphics{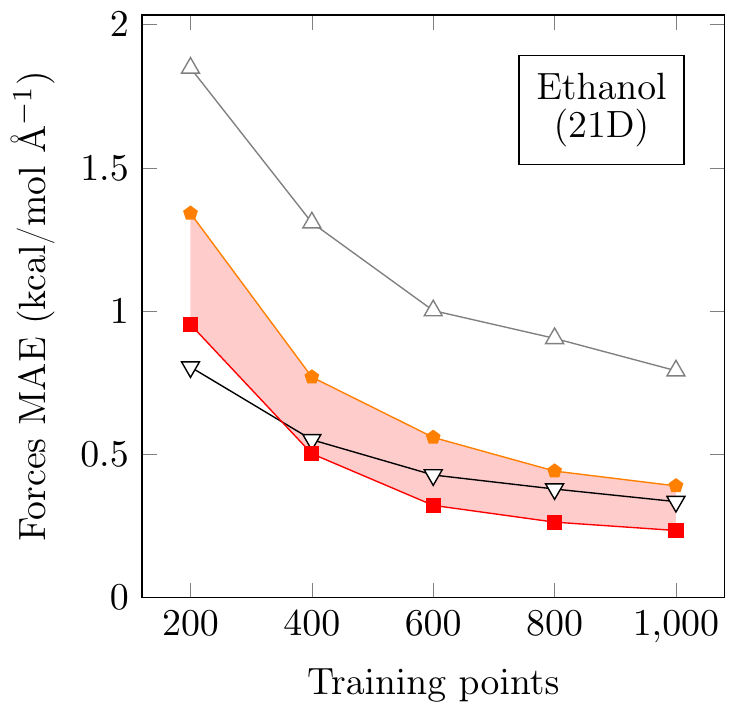} \\ 
		\end{tabular}
		\caption{Test energy and forces MAEs for GDML \cite{gdml-1} (up triangles) , sGDML \cite{gdml-2} (down triangles), \agdmlf (diamonds) and \agdmlcf (squares). The model abbreviations are defined in Table \ref{table:abbrev}.  The shaded area between the curves indicates the effect of increasing kernel complexity. The test errors are computed with 2000 molecular geometries.}
		\label{fig:comp_aniso_eth}
	\end{figure}

\begin{figure}[htbp]
		\begin{tabular}{cc}
			\includegraphics{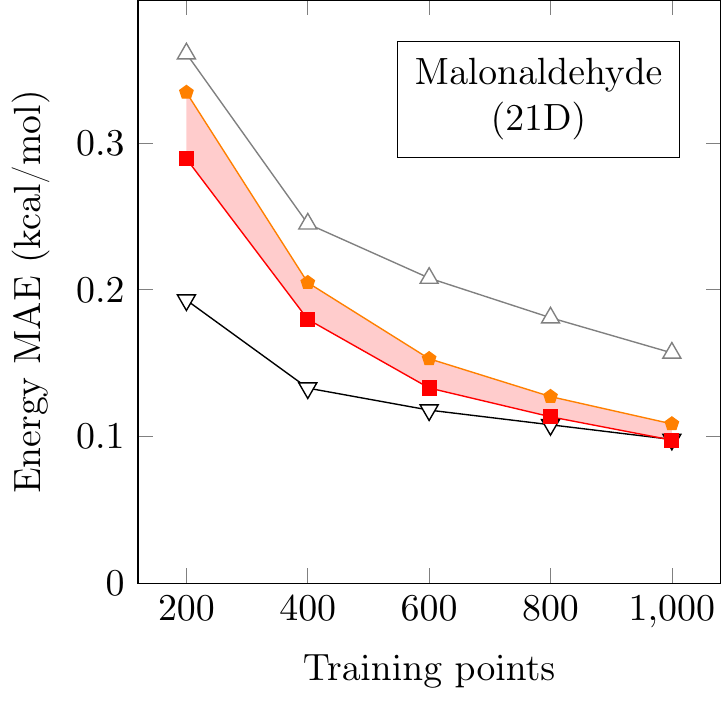} & \includegraphics{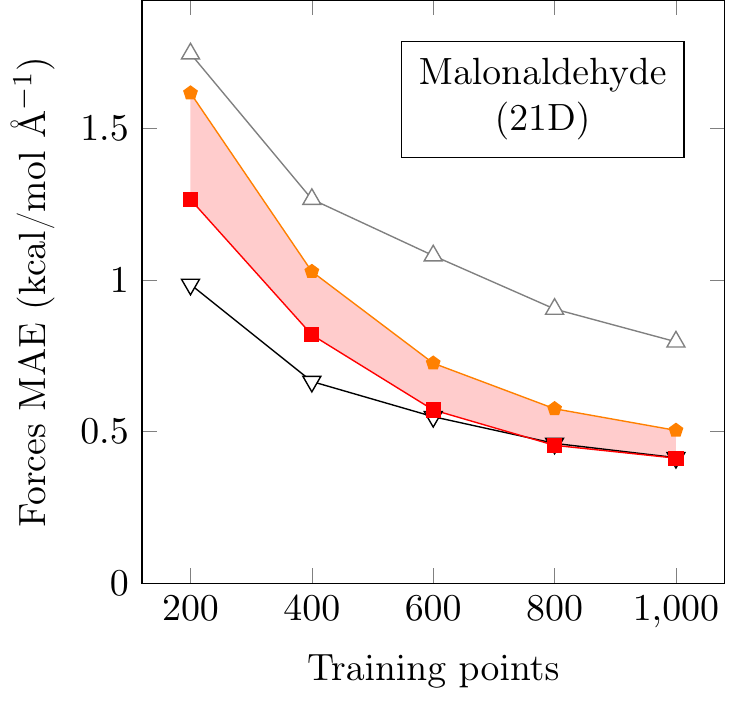} \\
		\end{tabular}
		\caption{Test energy and forces MAEs for GDML \cite{gdml-1} (up triangles) , sGDML \cite{gdml-2} (down triangles), \agdmlf (diamonds) and \agdmlcf (squares). The model abbreviations are defined in Table \ref{table:abbrev}.  The shaded area between the curves indicates the effect of increasing kernel complexity. The test errors are computed with 2000 molecular geometries.}
		\label{fig:comp_aniso_mal}
	\end{figure}

\begin{figure}[htbp]
		\begin{tabular}{cc}
			\includegraphics{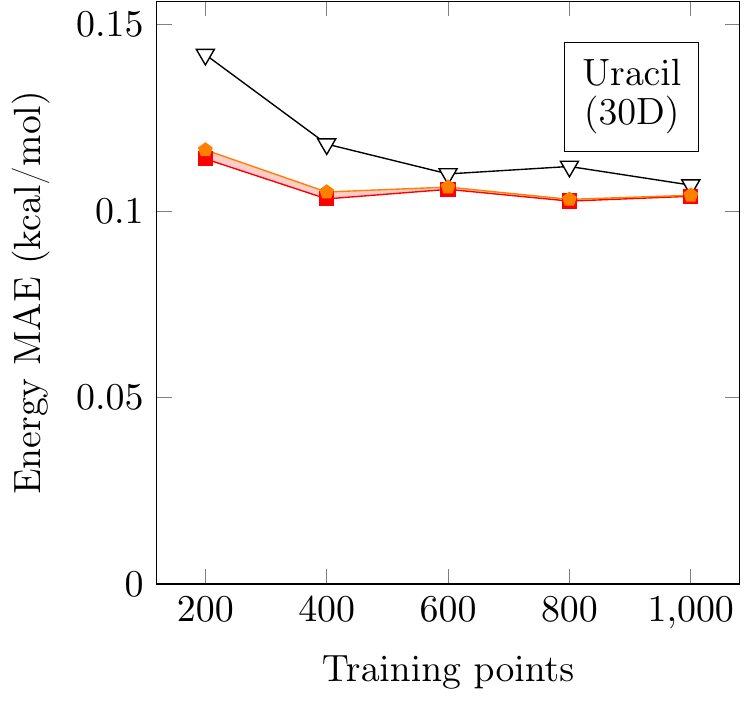} & \includegraphics{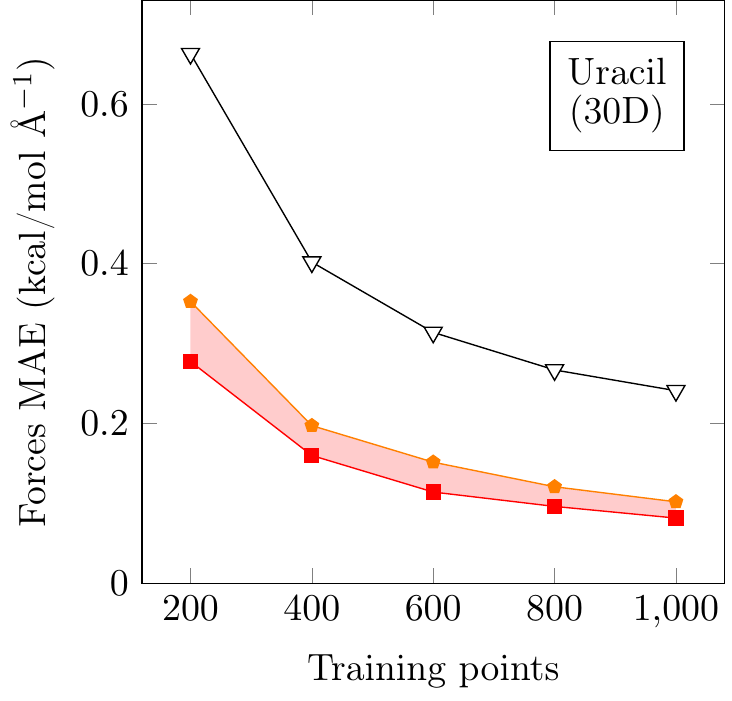} \\
		\end{tabular}
		\caption{Test energy and forces MAEs for GDML \cite{gdml-1} and sGDML \cite{gdml-2} (down triangles), \agdmlf (diamonds) and \agdmlcf (squares). The model abbreviations are defined in Table \ref{table:abbrev}.  The shaded area between the curves indicates the effect of increasing kernel complexity. The test errors are computed with 2000 molecular geometries.}
		\label{fig:comp_aniso_ura}
	\end{figure}

\begin{figure}[htbp]
		\begin{tabular}{cc}
			\includegraphics{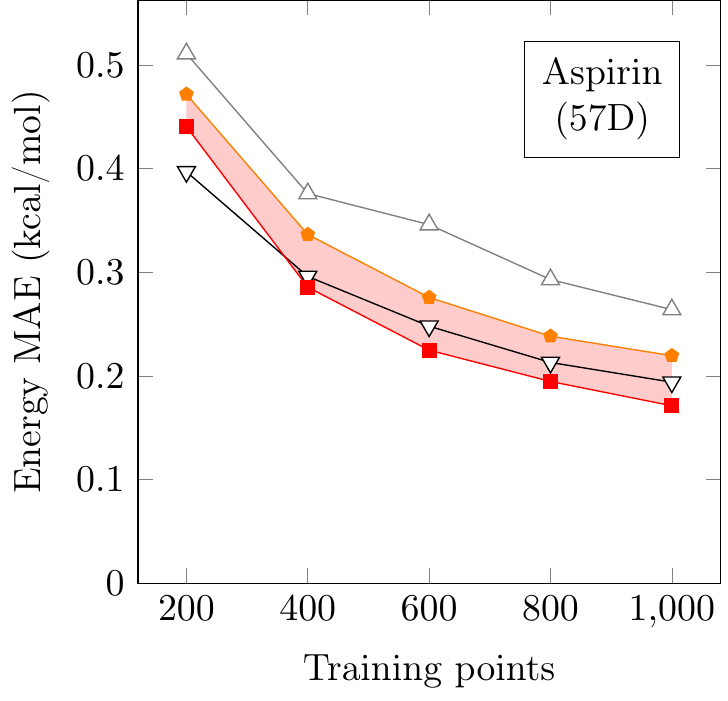} & \includegraphics{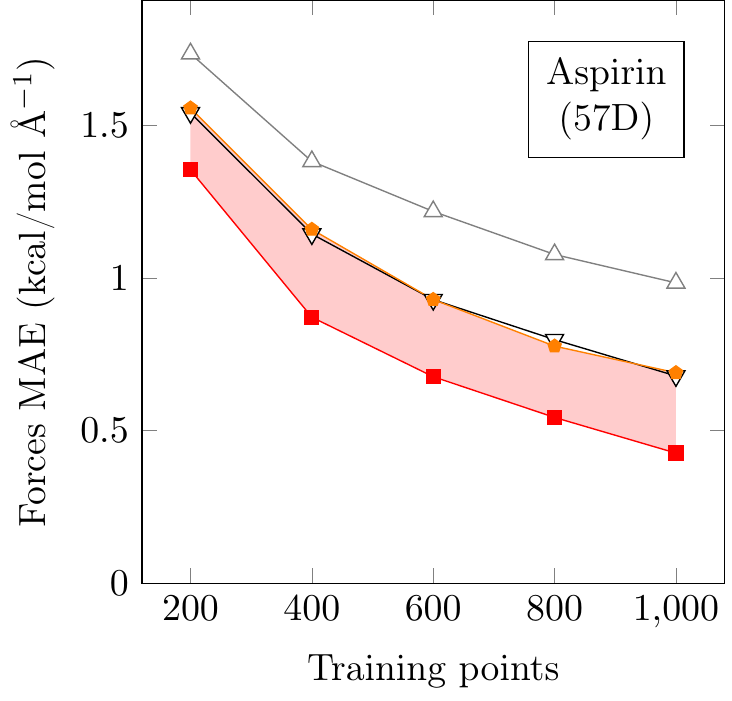} \\
		\end{tabular}
		\caption{Test energy and forces MAEs for GDML \cite{gdml-1} (up triangles) , sGDML \cite{gdml-2} (down triangles), \agdmlf (diamonds) and \agdmlcf (squares). The model abbreviations are defined in Table \ref{table:abbrev}.  The shaded area between the curves indicates the effect of increasing kernel complexity. The test errors are computed with 2000 molecular geometries.}
		\label{fig:comp_aniso_asp}
	\end{figure}

	Figures \ref{fig:comp_aniso_eth}, \ref{fig:comp_aniso_mal}, \ref{fig:comp_aniso_ura}, and \ref{fig:comp_aniso_asp} and Tables \ref{table:comp_aniso_eth}, \ref{table:comp_aniso_mal}, \ref{table:comp_aniso_ura} and \ref{table:comp_aniso_asp} display test errors of these models for predictions of both energies and forces for ethanol, malonaldehyde, uracil, and aspirin. In each case, the kernel functions of the \agdmlf model are chosen based on the values of BIC. 
	The open triangles show the previous results obtained with isotropic Mat\'ern 5/2 kernel functions: the up-triangles represent the GDML results and the down-triangles -- the sGDML results. 
	The shaded areas in Figures \ref{fig:comp_aniso_eth}, \ref{fig:comp_aniso_mal}, \ref{fig:comp_aniso_ura}, and \ref{fig:comp_aniso_asp}  show the improvement of the AGDML model accuracy due to increasing the complexity of the kernel functions. 
	The upper edges of the bands (diamonds) depict the AGDML results with simple kernels, whereas the lower edges (squares) correspond to the AGDML results with composite kernels.
	As uracil does not have any permutational symmetry, the reference GDML and sGDML models in Figure \ref{fig:comp_aniso_ura} have the same test error values and are displayed together as a single curve.
	We observe that the improvement due to the increasing complexity of GDML kernels is more significant for the prediction of forces and for larger molecules. 
	
	As evident from Figures \ref{fig:comp_aniso_eth}, \ref{fig:comp_aniso_mal}, \ref{fig:comp_aniso_ura}, \ref{fig:comp_aniso_asp}, the accuracy of the \agdmlcf models is comparable with and, in some cases, better  than the accuracy of the sGDML models. 
	Building symmetries into kernel functions is a molecule-specific task. In contrast, the kernel construction algorithm presented here is completely general and easy to automate. In addition, the method illustrated here can be applied to sGDML calculations, further improving the sGDML results. 
	
	As indicated by previous work \cite{gp-for-PES-12} with low dimensional molecules, the improvement due to composite kernels is more pronounced for smaller numbers of training points. However, models of large molecules with anisotropic kernels require a large number of kernel parameters. We observe that even for large molecules, models with composite kernels trained by a small number of training geometries are stable and generalize well. 
	An interesting case to examine is the model of aspirin with anisotropic composite kernels. The number of training parameters added at each layer of the kernel construction algorithm of Section \ref{composite_kernels} for the \agdmlcf model for aspirin is $\approx 210$. 
	This leads to models with $\approx 420$ trainable parameters for kernels with $2$ layers of complexity and $\approx 630$ trainable parameters for kernels with $3$ layers of complexity used to obtain results in Figure \ref{fig:comp_aniso_asp}.  In such models, the number of composite kernel parameters can be larger than the number of training points. 		
	This is not unusual in ML and occurs commonly in models based on complex neural networks. While such over-parameterized models are ill-defined from a statistical point of view, empirically they generalize well \cite{ml-params}. 
	We observe that models of aspirin based on composite kernels with $\approx 630$ parameters generalize better than models with simple anisotropic kernels, even when trained by $\leq 600$ training geometries.  
	
	\section{Conclusions}
	
	The generalization accuracy of ML models of PES and FFs for large polyatomic molecules can be typically improved by: (i) increasing the number of training points; (ii) improving the descriptors; (iii) improving the model. In order to build accurate models based on {\it ab initio} calculations, much of recent work has focused on (ii) and (iii). In particular, it has been shown that gradient domain machine learning models produce accurate results for high-dimensional molecular systems with a small number of {\it ab initio} calculations. The present work illustrates that GDML models can be further improved by increasing the complexity of underlying kernels through a greedy search algorithm with Bayesian information criterion as model selection. We have shown that this requires allowing for kernel anisotropy and produces significantly improved results. For example, we show that the GDML model trained by 1000 {\it ab initio} calculations produces a global 57-dimensional PES for aspirin with the mean absolute accuracy 0.177 kcal/mol (61.9 cm$^{-1}$) and FFs with the mean absolute accuracy 0.457 kcal/mol~\AA$^{-1}$. This requires composite kernels with 424 trainable parameters. We emphasize that such kernels cannot be chosen at random. As illustrated in Ref. \cite{gp-quant}, the iterative kernel construction algorithm based on BIC optimization is critical for building  models with low generalization error. 
	
	We note that the method demonstrated here can be used with either GDML or sGDML models. sGDML models take advantage of molecular symmetries to reduce the size of the input space, which results in better accuracy with the same number of training points. 
	The models presented here could also be improved 	by expanding the set of simple kernels to include more functions. This can increase the space of kernels, potentially offering more flexibility for the kernel optimization algorithm. In the present work, we have only included kernel functions that are doubly differentiable. It would be interesting to explore an approach that is based on combinations of doubly and singly differentiable kernel functions. 
	An example of a suitable singly differentiable function is a Mat\'ern 3/2 function. 
	The latter could be used to improve the accuracy of the PES models without affecting the models of FFs. The kernel optimization algorithm could further be improved by including products as well as linear combinations of kernels. This would complicate the computation of the Hessians but is not a fundamental obstacle. Finally, it would be interesting to include non-stationary kernels into the set of basis kernel functions. Non-stationary kernels have been proven important for extrapolation problems explored in previous studies \cite{gp-for-PES-11, extrapolation-2}. We have not studied the extrapolation properties of AGDML(c) models. Extrapolation outside the coordinate space of the training data is unlikely to produce accurate results with the current set of kernel functions that are all stationary. Previous studies using GPs for extrapolation \cite{gp-for-PES-11, extrapolation-2} have shown promising results predicting energies outside the range of energies in the training data. We expect GDML and AGDML(c) to also be effective at extrapolating in the energy or forces domain, but more in depth studies are needed.
	
	Our present work and the potential for further improvements thus indicate that it is possible to construct high-dimensional PES and FFs for large polyatomic molecules  with accuracy exceeding $0.1$ kcal/mol with a remarkably small number of quantum chemistry calculations (< 1000 for molecules with > 50 degrees of freedom). The kernel construction algorithm of the present work does not use any prior knowledge of the PES or FF landscape. The models are trained by forces at randomly chosen molecular geometries. There is no need for sophisticated sampling schemes such as ones based on active learning that usually require a significantly larger number of {\it ab initio} calculations than random sampling. The present approach is therefore readily applicable to any molecular systems. 
	
	\section{Acknowledgments}
	
	This work was supported by NSERC of Canada. 
	
	\section{Data Availability Statement}
	
	The data that support the findings of this study are openly available at the following URL: \href{http://quantum-machine.org/gdml/}{http://quantum-machine.org/gdml/}.

	\clearpage

	\clearpage
	\newpage
	
	\clearpage
	\newpage
	
	\begin{table}[htbp]
		\begin{NiceTabular}{|c|c|c|c|c|c|c|c|}[colortbl-like] \hline 
			\multirow{2}{*}{Molecule} & \multirow{2}{*}{\shortstack{Training \\ geometries}} & \multirow{2}{*}{Model type} & \multirow{2}{*}{\shortstack{Kernel \\ function}} & \multirow{2}{*}{\shortstack{Training \\ method}} & Test energy MAE & Test forces MAE & \multirow{2}{*}{Ref.} \\
			& & & & & (kcal/mol) & (kcal/mol~\AA$^{-1}$) & \\ \cline{6-8}
			\hline
			\multirow{8}{*}{\shortstack{Ethanol \\ (21D)}} & \multirow{8}{*}{400} & \multirow{6}{*}{\gdmlf} & \multirow{2}{*}{Mat\'ern 5/2} & LML & 0.280 & 1.385 & \multirow{6}{*}{This work} \\ \cline{5-7}
			& & & & Train E & 0.260 & 1.312 & \\ \cline{4-7}
			& & & \multirow{2}{*}{RBF} & LML & 0.256 & 1.401 & \\ \cline{5-7}
			& & & & Train E & 0.225 & 1.191 & \\ \cline{4-7}
			& & & \multirow{2}{*}{RatQuad} & LML & 0.237 & 1.314 & \\\cline{5-7}
			& & & & Train E & \cellcolor{red!25}\textbf{0.225} & \cellcolor{red!25}\textbf{1.191} & \\ \cline{3-8}
			& & GDML & Mat\'ern 5/2 & Grid search & 0.261 & 1.309 & \cite{gdml-1} \\ \cline{3-8}
			& & sGDML & Mat\'ern 5/2 & Grid search & 0.103 & 0.551 & \cite{gdml-2} \\ 
			\hline
		\end{NiceTabular}
		\caption{Comparison of GDML models with isotropic simple kernels ($\lambda = 10^{-10}$) trained using log marginal likelihood and training energies with errors computed on 2000 test geometries with the previously published GDML \cite{gdml-1} and sGDML \cite{gdml-2} results. Best performing GDML models are highlighted in red.}
		\label{table:simple_iso}
	\end{table}
	
	
	\begin{table}[htbp]
		\begin{NiceTabular}{|c|c|c|c|c|c|c|c|}[colortbl-like] \hline 
			\multirow{2}{*}{Molecule} & \multirow{2}{*}{\shortstack{Training \\ geometries}} & \multirow{2}{*}{Model type} & \multirow{2}{*}{\shortstack{Kernel \\ function}} & \multirow{2}{*}{\shortstack{Training \\ method}} & Test energy MAE & Test forces MAE & \multirow{2}{*}{Ref.} \\
			& & & & & (kcal/mol) & (kcal/mol~\AA$^{-1}$) & \\ \cline{6-8}
			\hline
			\multirow{9}{*}{\shortstack{Ethanol \\ (21D)}} & \multirow{9}{*}{400} & \multirow{6}{*}{\agdmlf} & \multirow{2}{*}{Mat\'ern 5/2} & LML & 0.188 & 0.923 & \multirow{7}{*}{This work} \\ \cline{5-7}
			& & & & Train E & 0.220 & 1.144 & \\ \cline{4-7}
			& & & \multirow{2}{*}{RBF} & LML & 0.148 & 0.775 & \\ \cline{5-7}
			& & & & Train E & 0.187 & 1.010 & \\ \cline{4-7}
			& & & \multirow{2}{*}{RatQuad} & LML & \cellcolor{red!25}\textbf{0.147} & \cellcolor{red!25}\textbf{0.770} & \\\cline{5-7}
			& & & & Train E & 0.188 & 1.034 & \\ \cline{3-7}
			& & \gdmlf & RatQuad & Train E & \cellcolor{blue!25}\textbf{0.225} & \cellcolor{blue!25}\textbf{1.191} & \\ \cline{3-8}
			& & GDML\cite{gdml-1} & Mat\'ern 5/2 & Grid search & 0.261 & 1.309 & \cite{gdml-1} \\ \cline{3-8}
			& & sGDML\cite{gdml-2} & Mat\'ern 5/2 & Grid search & 0.103 & 0.551 & \cite{gdml-2} \\ 
			\hline
		\end{NiceTabular}
		\caption{Comparison of GDML models with anisotropic simple kernels ($\lambda = 10^{-10}$) trained using log marginal likelihood and training energies with errors computed on 2000 test geometries with the previously published GDML \cite{gdml-1} and sGDML \cite{gdml-2} results. Best performing GDML models are highlighted in red. The best results with simple isotropic kernels from Table \ref{table:simple_iso} are highlighted in blue. }
		\label{table:simple_aniso}
	\end{table}

	\begin{table}[htbp]
		\begin{NiceTabular}{|c|c|c|c|c|c|}[colortbl-like] \hline 
			\multirow{2}{*}{Molecule} & \multirow{2}{*}{\shortstack{Training \\ geometries}} & \multirow{2}{*}{Model} & Test energy MAE & Test forces MAE & \multirow{2}{*}{Ref.} \\
			& & & (kcal/mol) & (kcal/mol~\AA$^{-1}$) \\
			\hline
			\multirow{20}{*}{\shortstack{Ethanol \\ (21D)}} & \multirow{4}{*}{200} & \agdmlcf & \multirow{2}{*}{0.196} & \multirow{2}{*}{0.954} & \multirow{2}{*}{This work} \\ 
			& & (4 layers) & & & \\ \cline{3-6}
			& & GDML & 0.392 & 1.849 & \cite{gdml-1} \\ \cline{3-6}
			& & sGDML & \cellcolor{red!25}\textbf{0.151} & \cellcolor{red!25}\textbf{0.805} & \cite{gdml-2} \\ \cline{2-6}
			& \multirow{4}{*}{400} & \agdmlcf & \multirow{2}{*}{0.108} & \multirow{2}{*}{\cellcolor{red!25}\textbf{0.502}} & \multirow{2}{*}{This work} \\ 
			& & (4 layers) & & \cellcolor{red!25} & \\ \cline{3-6}
			& & GDML & 0.261 & 1.309 & \cite{gdml-1} \\ \cline{3-6}
			& & sGDML & \cellcolor{red!25}\textbf{0.103} & 0.551 & \cite{gdml-2} \\ \cline{2-6}
			& \multirow{4}{*}{600} & \agdmlcf & \cellcolor{red!25}\textbf{\multirow{2}{*}{0.0792}} & \cellcolor{red!25}\textbf{\multirow{2}{*}{0.322}} & \multirow{2}{*}{This work} \\ 
			& & (4 layers) & \cellcolor{red!25} & \cellcolor{red!25} & \\ \cline{3-6}
			& & GDML & 0.194 & 1.002 & \cite{gdml-1} \\ \cline{3-6}
			& & sGDML & 0.083 & 0.428 & \cite{gdml-2} \\ \cline{2-6}
			& \multirow{4}{*}{800} & \agdmlcf & \cellcolor{red!25}\textbf{\multirow{2}{*}{0.0692}} & \cellcolor{red!25}\textbf{\multirow{2}{*}{0.263}} & \multirow{2}{*}{This work} \\ 
			& & (4 layers) & \cellcolor{red!25} & \cellcolor{red!25} & \\ \cline{3-6}
			& & GDML & 0.171 & 0.905 & \cite{gdml-1} \\ \cline{3-6}
			& & sGDML & 0.077 & 0.379 & \cite{gdml-2} \\ \cline{2-6}
			& \multirow{4}{*}{1000} & \agdmlcf & \cellcolor{red!25}\textbf{\multirow{2}{*}{0.0653}} & \cellcolor{red!25}\textbf{\multirow{2}{*}{0.234}} & \multirow{2}{*}{This work} \\ 
			& & (4 layers) & \cellcolor{red!25} & \cellcolor{red!25} & \\ \cline{3-6}
			& & GDML & 0.154 & 0.792 & \cite{gdml-1} \\ \cline{3-6}
			& & sGDML & 0.072 & 0.335 & \cite{gdml-2} \\ 
			\hline
		\end{NiceTabular}
		\caption{Comparison of GDML models with anisotropic composite kernels ($\lambda = 10^{-6}$) trained using log marginal likelihood with errors computed on 2000 test geometries with the previously published GDML \cite{gdml-1} and sGDML \cite{gdml-2} results. The lowest error values for each number of training geometries is highlighted in red.}
		\label{table:comp_aniso_eth}
	\end{table}
	
	\begin{table}[htbp]
		\begin{NiceTabular}{|c|c|c|c|c|c|}[colortbl-like] \hline 
			\multirow{2}{*}{Molecule} & \multirow{2}{*}{\shortstack{Training \\ geometries}} & \multirow{2}{*}{Model} & Test energy MAE & Test forces MAE & \multirow{2}{*}{Ref.} \\
			& & & (kcal/mol) & (kcal/mol~\AA$^{-1}$) \\
			\hline
			\multirow{20}{*}{\shortstack{Malonaldehyde \\ (21D)}} & \multirow{4}{*}{200} & \agdmlcf & \multirow{2}{*}{0.290} & \multirow{2}{*}{1.265} & \multirow{2}{*}{This work} \\ 
			& & (3 layers) & & & \\ \cline{3-6}
			& & GDML & 0.361 & 1.746 & \cite{gdml-1} \\ \cline{3-6}
			& & sGDML & \cellcolor{red!25}\textbf{0.193} & \cellcolor{red!25}\textbf{0.985} & \cite{gdml-2} \\ \cline{2-6}
			& \multirow{4}{*}{400} & \agdmlcf & \multirow{2}{*}{0.180} & \multirow{2}{*}{0.819} & \multirow{2}{*}{This work} \\ 
			& & (3 layers) & & & \\ \cline{3-6}
			& & GDML & 0.245 & 1.266 & \cite{gdml-1} \\ \cline{3-6}
			& & sGDML & \cellcolor{red!25}\textbf{0.133} & \cellcolor{red!25}\textbf{0.665} & \cite{gdml-2} \\ \cline{2-6}
			& \multirow{4}{*}{600} & \agdmlcf & \multirow{2}{*}{0.133} & \multirow{2}{*}{0.571} & \multirow{2}{*}{This work} \\ 
			& & (3 layers) & & & \\ \cline{3-6}
			& & GDML & 0.208 & 1.080 & \cite{gdml-1} \\ \cline{3-6}
			& & sGDML & \cellcolor{red!25}\textbf{0.118} & \cellcolor{red!25}\textbf{0.549} & \cite{gdml-2} \\ \cline{2-6}
			& \multirow{4}{*}{800} & \agdmlcf & \multirow{2}{*}{0.113} & \multirow{2}{*}{\cellcolor{red!25}\textbf{0.454}} & \multirow{2}{*}{This work} \\ 
			& & (3 layers) & & \cellcolor{red!25} & \\ \cline{3-6}
			& & GDML & 0.181 & 0.904 & \cite{gdml-1} \\ \cline{3-6}
			& & sGDML & \cellcolor{red!25}\textbf{0.108} & 0.461 & \cite{gdml-2} \\ \cline{2-6}
			& \multirow{4}{*}{1000} & \agdmlcf & \multirow{2}{*}{\cellcolor{red!25}\textbf{0.098}} & \multirow{2}{*}{\cellcolor{red!25}\textbf{0.389}} & \multirow{2}{*}{This work} \\ 
			& & (3 layers) & \cellcolor{red!25} & \cellcolor{red!25} & \\ \cline{3-6}
			& & GDML & 0.157 & 0.796 & \cite{gdml-1} \\ \cline{3-6}
			& & sGDML & 0.098 & 0.414 & \cite{gdml-2} \\ 
			\hline
		\end{NiceTabular}
		\caption{Comparison of GDML models with anisotropic composite kernels ($\lambda = 10^{-6}$) trained using log marginal likelihood with errors computed on 2000 test geometries with the previously published GDML \cite{gdml-1} and sGDML \cite{gdml-2} results. The lowest error values for each number of training geometries is highlighted in red. }
		\label{table:comp_aniso_mal}
	\end{table}
	
	\begin{table}[htbp]
		\begin{NiceTabular}{|c|c|c|c|c|c|}[colortbl-like] \hline 
			\multirow{2}{*}{Molecule} & \multirow{2}{*}{\shortstack{Training \\ geometries}} & \multirow{2}{*}{Model} & Test energy MAE & Test forces MAE & \multirow{2}{*}{Ref.} \\
			& & & (kcal/mol) & (kcal/mol~\AA$^{-1}$) \\
			\hline
			\multirow{15}{*}{\shortstack{Uracil \\ (30D)}} & \multirow{3}{*}{200} & \agdmlcf & \multirow{2}{*}{\cellcolor{red!25}\textbf{0.114}} & \multirow{2}{*}{\cellcolor{red!25}\textbf{0.278}} & \multirow{2}{*}{This work} \\ 
			& & (4 layers) & \cellcolor{red!25} & \cellcolor{red!25} & \\ \cline{3-6}
			& & (s)GDML & 0.142 & 0.663 & \cite{gdml-1}, \cite{gdml-2} \\ \cline{2-6}
			& \multirow{3}{*}{400} & \agdmlcf & \multirow{2}{*}{\cellcolor{red!25}\textbf{0.103}} & \multirow{2}{*}{\cellcolor{red!25}\textbf{0.163}} & \multirow{2}{*}{This work} \\ 
			& & (4 layers) & \cellcolor{red!25} & \cellcolor{red!25} & \\ \cline{3-6}
			& & (s)GDML & 0.118 & 0.402 & \cite{gdml-1}, \cite{gdml-2} \\ \cline{2-6}
			& \multirow{3}{*}{600} & \agdmlcf & \multirow{2}{*}{\cellcolor{red!25}\textbf{0.106}} & \multirow{2}{*}{\cellcolor{red!25}\textbf{0.114}} & \multirow{2}{*}{This work} \\ 
			& & (4 layers) & \cellcolor{red!25} & \cellcolor{red!25} & \\ \cline{3-6}
			& & (s)GDML & 0.110 & 0.314 & \cite{gdml-1}, \cite{gdml-2} \\ \cline{2-6}
			& \multirow{3}{*}{800} & \agdmlcf & \multirow{2}{*}{\cellcolor{red!25}\textbf{0.103}} & \multirow{2}{*}{\cellcolor{red!25}\textbf{0.096}} & \multirow{2}{*}{This work} \\ 
			& & (3 layers) & \cellcolor{red!25} & \cellcolor{red!25} & \\ \cline{3-6}
			& & (s)GDML & 0.112 & 0.267 & \cite{gdml-1}, \cite{gdml-2} \\ \cline{2-6}
			& \multirow{3}{*}{1000} & \agdmlcf & \multirow{2}{*}{\cellcolor{red!25}\textbf{0.104}} & \multirow{2}{*}{\cellcolor{red!25}\textbf{0.082}} & \multirow{2}{*}{This work} \\ 
			& & (3 layers) & \cellcolor{red!25} & \cellcolor{red!25} & \\ \cline{3-6}
			& & (s)GDML & 0.107 & 0.241 & \cite{gdml-1}, \cite{gdml-2} \\ 
			\hline
		\end{NiceTabular}
		\caption{Comparison of GDML models with anisotropic composite kernels ($\lambda = 10^{-6}$) trained using log marginal likelihood with errors computed on 2000 test geometries with the previously published GDML \cite{gdml-1} and sGDML \cite{gdml-2} results. The lowest error values for each number of training geometries is highlighted in red.}
		\label{table:comp_aniso_ura}
	\end{table}
	
	\begin{table}[htbp]
		\begin{NiceTabular}{|c|c|c|c|c|c|}[colortbl-like] \hline 
			\multirow{2}{*}{Molecule} & \multirow{2}{*}{\shortstack{Training \\ geometries}} & \multirow{2}{*}{Model} & Test energy MAE & Test forces MAE & \multirow{2}{*}{Ref.} \\
			& & & (kcal/mol) & (kcal/mol~\AA$^{-1}$) \\
			\hline
			\multirow{20}{*}{\shortstack{Aspirin \\ (57D)}} & \multirow{4}{*}{200} & \agdmlcf & \multirow{2}{*}{0.441} & \multirow{2}{*}{\cellcolor{red!25}\textbf{1.355}} & \multirow{2}{*}{This work} \\ 
			& & (3 layers) & & \cellcolor{red!25} & \\ \cline{3-6}
			& & GDML & 0.511 & 1.735 & \cite{gdml-1} \\ \cline{3-6}
			& & sGDML & \cellcolor{red!25}\textbf{0.397} & 1.541 & \cite{gdml-2} \\ \cline{2-6}
			& \multirow{4}{*}{400} & \agdmlcf & \multirow{2}{*}{\cellcolor{red!25}\textbf{0.286}} & \multirow{2}{*}{\cellcolor{red!25}\textbf{0.872}} & \multirow{2}{*}{This work} \\ 
			& & (3 layers) & \cellcolor{red!25} & \cellcolor{red!25} & \\ \cline{3-6}
			& & GDML & 0.376 & 1.382 & \cite{gdml-1} \\ \cline{3-6}
			& & sGDML & 0.296 & 1.144 & \cite{gdml-2} \\ \cline{2-6}
			& \multirow{4}{*}{600} & \agdmlcf & \multirow{2}{*}{\cellcolor{red!25}\textbf{0.225}} & \multirow{2}{*}{\cellcolor{red!25}\textbf{0.677}} & \multirow{2}{*}{This work} \\ 
			& & (2 layers) & \cellcolor{red!25} & \cellcolor{red!25} & \\ \cline{3-6}
			& & GDML & 0.346 & 1.218 & \cite{gdml-1} \\ \cline{3-6}
			& & sGDML & 0.248 & 0.929 & \cite{gdml-2} \\ \cline{2-6}
			& \multirow{4}{*}{800} & \agdmlcf & \multirow{2}{*}{\cellcolor{red!25}\textbf{0.195}} & \multirow{2}{*}{\cellcolor{red!25}\textbf{0.543}} & \multirow{2}{*}{This work} \\ 
			& & (2 layers) & \cellcolor{red!25} & \cellcolor{red!25} & \\ \cline{3-6}
			& & GDML & 0.293 & 1.077 & \cite{gdml-1} \\ \cline{3-6}
			& & sGDML & 0.213 & 0.798 & \cite{gdml-2} \\ \cline{2-6}
			& \multirow{4}{*}{1000} & \agdmlcf & \multirow{2}{*}{\cellcolor{red!25}\textbf{0.177}} & \multirow{2}{*}{\cellcolor{red!25}\textbf{0.457}} & \multirow{2}{*}{This work} \\ 
			& & (2 layers) & \cellcolor{red!25} & \cellcolor{red!25} & \\ \cline{3-6}
			& & GDML & 0.264 & 0.984 & \cite{gdml-1} \\ \cline{3-6}
			& & sGDML & 0.194 & 0.679 & \cite{gdml-2} \\ 
			\hline
		\end{NiceTabular}
		\caption{Comparison of GDML models with anisotropic composite kernels ($\lambda = 10^{-6}$) trained using log marginal likelihood with errors computed on 2000 test geometries with the previously published GDML \cite{gdml-1} and sGDML \cite{gdml-2} results. The lowest error values for each number of training geometries is highlighted in red.}
		\label{table:comp_aniso_asp}
	\end{table}

	\clearpage
	\newpage
	
	\nocite{*}
	\bibliography{manuscript}
	
\end{document}